\documentclass[fleqn,usenatbib]{mnras}

\usepackage{newtxtext,newtxmath}

\usepackage[T1]{fontenc}
\usepackage{ae,aecompl}

\usepackage[utf8]{inputenc}
\usepackage{graphicx}	

\newcommand{\corr}[1]{{#1}}

\title[Coronagraphic phase diversity through residual turbulence]{Coronagraphic phase diversity through residual turbulence: performance study and experimental validation}

\author[O. Herscovici-Schiller, J.-F. Sauvage, L. M. Mugnier, K. Dohlen \& A. Vigan]{
Olivier Herscovici-Schiller,$^{1}$\thanks{E-mail: olivier.herscovici@onera.fr (OHS)}
Jean-François Sauvage,$^{2,3}$
Laurent M. Mugnier,$^{2}$ 
\newauthor
Kjetil Dohlen$^{3}$ and Arthur Vigan$^{3}$
\\
$^{1}$ DTIS, ONERA, Université Paris Saclay, F-91123 Palaiseau -- France\\
$^{2}$ DOTA, ONERA, Université Paris Saclay, F-92322 Châtillon -- France\\
$^{3}$ Laboratoire d’Astrophysique de Marseille UMR 7326, Aix-Marseille Université, CNRS, 13388 Marseille, France\\
}

\date{Accepted 2019 July 15. Received 2019 July 15; in original form 2019 January 17}

\pubyear{2019}

\begin{document}
\label{firstpage}
\pagerange{\pageref{firstpage}--\pageref{lastpage}}
\maketitle

\begin{abstract}
Quasi-static aberrations in coronagraphic systems are the ultimate limitation to the capabilities of exoplanet imagers both ground-based and space-based. These aberrations -- which can be due to various causes such as optics alignment or moving optical parts during the observing sequence -- create light residuals called speckles in the focal plane that might be mistaken for a planets. For ground-based instruments, the presence of residual turbulent wavefront errors due to partial adaptive optics correction causes an additional difficulty to the challenge of measuring aberrations in the presence of a coronagraph. In this paper, we present an extension of COFFEE, the coronagraphic phase diversity, to the estimation of quasi-static aberrations in the presence of adaptive optics-corrected residual turbulence. We perform realistic numerical simulations to assess the performance that can be expected on an instrument of the current generation. We perform the first experimental validation in the laboratory which demonstrates that quasi-static aberrations can be corrected during the observations by means of coronagraphic phase diversity.
\end{abstract}




\section{Introduction}
    Imaging exoplanets is a challenging task. Current ground-based exoplanet imagers such as SPHERE or GPI reach contrast levels of $10^{-6}$ in H-band at 0.5 arc second separations~\citep{beuzit2008sphere,macintosh2008gemini}. Such high contrast imaging is enabled by the use of coronagraphs, devices which reject the on-axis starlight but let out-of-axis light from disks or planets pass through. However, any optical aberration in the instrument causes light to leak through the coronagraph, which results in speckles appearing in the focal plane of the telescope. These speckles constitute the ultimate contrast limit for exoplanet imagers. 

    Since the quasi-static aberrations evolve slowly during the night \citep{martinez_speckle_2012,martinez_speckle_2013}, the full contrast capacity of exoplanet imagers can be reached only if a correction is applied during the night. A few methods have been developed to measure the quasi-static aberrations~\citep{ndiaye_calibration_2013,galicher_wavefront_2008}. However, none of them is currently adapted to being used during the observations of a ground-based instrument, although some tests are currently taking place~\citep{vigan_-sky_2018}. Our goal is to present the adequacy of the coronagraphic phase diversity, COFFEE, to the measurement of quasi-static phase aberrations in the presence of adaptive-optics-corrected residual turbulence.
    We start with the adaptation of the algorithm. We continue with a performance assessment consisting in numerical studies of the sensitivity of the method to various sources of perturbation. We finish with the description of a laboratory validation of the coronagraphic phase diversity in the presence of residual turbulence on the MITHiC bench at laboratoire d’astrophysique de Marseille.
    
\section{Formalism of the coronagraphic phase diversity in the presence of residual adaptive-optics-corrected turbulence}
\label{sec:coffee}
    In this section, we recall the formalism of coronagraphic phase diversity and we present its adaptation to the presence of adaptive-optics-corrected turbulence.
    \subsection{COFFEE, the coronagraphic phase diversity}
        COFFEE, the coronagraphic phase diversity, is a post-coronagraphic wavefront sensor \citep{sauvage_coronagraphic_2012,paul_high-order_2013}. As its name suggests, it is a flavour of the phase diversity \citep{gonsalves_phase_1982,mugnier_phase_2006} method. The principle is to use information encoded in images produced by the scientific instrument to retrieve the aberrations of the instrument. Unfortunately, more than only one image is needed to do so. In phase diversity, two (or more) images are used. One is called the focused image. The other one, called the diversity image, is taken while a known aberration, called the diversity phase, is voluntarily introduced in the system. In the context of coronagraphic phase diversity for ground-based instruments, the diversity phase can be easily introduced by the deformable mirror of the adaptive optics system. 
        
        From a mathematical point of view, COFFEE is a \textit{maximum a posteriori} estimator with non-homogeneous Gaussian noise assumption. The principle is to find the upstream phase aberration \(\widehat{\phi_\mathrm{up}}\) and the downstream phase aberration \(\widehat{\phi_\mathrm{down}}\) that minimise the cost function \(\mathcal{J}\):
        \begin{equation}
            \mathcal{J}\left(\phi\right) = \sum_{k,x,y} \dfrac{\left\| i(k,x,y) - m(\phi_\mathrm{up},\phi_\mathrm{down},k,x,y) \right\|^2}{2 \sigma^2(k,x,y)} + \mathcal{R}(\phi_\mathrm{up}) + \mathcal{R}(\phi_\mathrm{down}) .
            \label{eq:coffee}
        \end{equation}
        In the first term of the right hand sign, $i$ is an experimental image produced by the detector. Its counterpart $m$ is a numerical model of the image, which takes into account the sought quasi-static aberrations $\phi_\mathrm{up}$ and $\phi_\mathrm{down}$. The indexes of the sum are $x$ and $y$, which are the coordinates of the pixels in images, and $k$, which is there to distinguish between the focal image and the diversity image. The denominator $\sigma^2$ is the variance of the noise level of pixel $x,y$ in the image $i(k)$. This first term is a noise-weighted distance between the experimental data and the output of a model. The second term of the right hand sign, $\mathcal{R}$, is a regularisation term, often taken as 
        \begin{equation}
            \mathcal{R}(\phi) = \dfrac{1}{2\sigma^2_{\nabla \phi}}\sum_{x,y}\left\|\nabla \phi \right\|^2(x,y),
            \label{eq:regul}
        \end{equation}
        where $\nabla\phi$ is the spatial gradient of $\phi$. This second term represents prior knowledge of the statistics of the aberrations. More precisely, this terms penalises the phase spatial gradients, which smooths the reconstructed phase and attenuates the noise propagation in the wavefront. This avoids unrealistic very high spatial frequencies to appear in the reconstructed wavefront. 
      
        \corr{If the images are not narrow-band, the impact of spectral width is negligible for $\Delta~\lambda/\lambda < 15\%$ \citep{meynadier1999noise}. If the image is broad-band, then the numerical model of the image must be calculated for several wavelengths~\citep{seldin2000closed}. This will increase the computation cost of the numerical model in proportion to the width of the spectral band. However, the calculations at different wavelengths can absolutely be done in parallel, so the increase in the number of computations will not result in an increase in the duration of the computation if a multiple-core-computer can be used. }
        
    \subsection{Implementation: taking turbulence into account}
        The numerical implementation of COFFEE in a turbulence-free case is described in \cite{paul_high-order_2013} and \cite{paul_coronagraphic_2013}.
        The expression of $m$ in COFFEE is
        \begin{equation}
            m(\phi,k,x,y) = f_k\times\left[h_\mathrm{det} \star h_\mathrm{c}(\phi+\phi_k)\right] (x,y) + b_k,
        \end{equation}
        where $f_k$ the incoming flux, $h_\mathrm{det}$ is the response of the detector, $\star$ is the convolution operator, $h_\mathrm{c}$ is the point spread function of the coronagraphic instrument, $\phi$ is the static aberration that we seek to retrieve, $\phi_k$ is zero for the focused image and is the diversity phase in the diversity image, and $b_k$ is a constant background.

        In order to take the effect of atmospheric turbulence into account, the model of data formation $m$ must reckon the impact of turbulence. In order to do so, we developed an analytic expression for coronagraphic imaging through turbulence \citep{herscovici-schiller_analytic_2017} to use as the point spread function of the instrument.
        The optical impulse response of the coronagraphic instrument in the presence of (residual) turbulence is noted $h_{\rm lec}$ -- the index stands for “long exposure coronagraphic”.  
        We still note $h_\mathrm{c}$ the impulse response of the coronagraphic instrument without turbulence. \corr{By analogy with the atmospheric transfer function \citep{roddier1981} as described by \cite{herscovici-schiller_analytic_2017},} we note $h_\mathrm{a}(D_\phi)$ the atmospheric point spread function, defined as 
        \begin{equation}
            h_\mathrm{a}(D_\phi) = \mathcal{F}^{-1}\left[ \exp\left(- \frac{1}{2}D_\phi\right)\right],
            \label{eq:ha}
        \end{equation}
        where $D_\phi$ is the phase structure function of the (residual) atmospheric turbulence. If the turbulence is supposed to be stationary and ergodic, then, for an exposure time much larger than the characteristic time of turbulence,
        \begin{align}
            h_\mathrm{lec}&(\alpha; \phi_\mathrm{up}, \phi_\mathrm{down}, D_\phi)  = \nonumber \\
            & \iint h_\mathrm{a}\left(\alpha'; D_\phi\right) \times h_\mathrm{c} \left( \alpha; \phi_\mathrm{up}+2\uppi \alpha'\cdot \mathrm{Id}, \phi_\mathrm{down} \right) \, \mathrm{d} \alpha'  ,
            \label{eq:im_corono_turb}
        \end{align}
        where $\alpha$ is a two-dimensional angular position in the focal plane, $\phi_\mathrm{up}$ is the static aberration upstream of the coronagraph, $\phi_\mathrm{down}$ is the static aberration downstream of the coronagraph, and Id is the identity function of $\mathbb{R}^2$. 
        In order to use this forward model into COFFEE, one needs to compute it efficiently, and to compute its gradients.

        The numeric model of the instrument is computed by performing a discrete sum to approximate the integral, using the same Fourier optics model for $h_\mathrm{c}$ as in \cite{paul_high-order_2013}, as long as an estimate of $D_\phi$ is available, for example using one of the methods described in~\cite{sauvage_coronagraphic_2012} or \cite{veran97}. If the point spread function is sampled on $N\times N$ pixels, a natural choice is to calculate the numeric point spread function as
        \begin{align}
            h_\mathrm{lec}&(\alpha; \phi_\mathrm{up}, \phi_\mathrm{down}, D_\phi)  =  \\
            & \sum_{\alpha'_x =-N/2}^{N/2}\sum_{\alpha'_y =-N/2}^{N/2} h_\mathrm{a}\left(\alpha'; D_\phi\right) \times h_\mathrm{c} \left( \alpha; \phi_\mathrm{up}+2\uppi \alpha'\cdot \mathrm{Id}, \phi_\mathrm{down} \right). \nonumber
        \end{align}
    
     \subsection{A physical approximation that reduces computing costs}
        The resulting cost of calculation is then $N^2$ times the cost of calculating $h_\mathrm{c}$. This cost can be considerably alleviated if the coronagraph is a Lyot-type coronagraph such as Lyot's opaque mask or Roddier's phase mask coronagraphs. 
        Indeed, for those coronagraphs consisting of small phase or amplitude features strictly located in the vicinity of the stellar image, the influence of the focal mask on the point spread function is essentially negligible when there is a strong upstream tip-tilt. Let us define a central square region of side $M$ in the focal plane of the coronagraphic mask. If the light beam is centred outside of this central square, the coronagraph is supposed to have no influence on the beam. 
        Then, the previous equation can be split into two regions:
        \begin{align}
         & h_\mathrm{lec}(\alpha; \phi_\mathrm{up}, \phi_\mathrm{down}, D_\phi)  =  \\
           &  \sum_{\substack{\alpha'_x =-N/2 \\ \alpha'_x \not \in [-M/2,M/2]}}^{N/2}\sum_{\substack{\alpha'_y =-N/2 \\ \alpha'_y \not \in [-M/2,M/2]}}^{N/2} h_\mathrm{a}\left(\alpha'; D_\phi\right) \times h_\mathrm{c} \left( \alpha; \phi_\mathrm{up}+2\uppi \alpha'\cdot \mathrm{Id}, \phi_\mathrm{down} \right) \nonumber \\
           & + \sum_{\alpha'_x =-M/2}^{M/2}\sum_{\alpha'_y =-M/2}^{M/2} h_\mathrm{a}\left(\alpha'; D_\phi\right) \times h_\mathrm{c} \left( \alpha; \phi_\mathrm{up}+2\uppi \alpha'\cdot \mathrm{Id}, \phi_\mathrm{down} \right). \nonumber
        \end{align}
        Since the double sum on the first line exists only for strong tip-tilts, the value of $h_\mathrm{c}$ in it is very close to the value of a non-coronagraphic point spread function, denoted $h$:
        \begin{align}
         & h_\mathrm{lec}(\alpha; \phi_\mathrm{up}, \phi_\mathrm{down}, D_\phi)  \approx \label{eq:inter_approx_convol} \\
           &  \sum_{\substack{\alpha'_x =-N/2 \\ \alpha'_x \not \in [-M/2,M/2]}}^{N/2}\sum_{\substack{\alpha'_y =-N/2 \\ \alpha'_y \not \in [-M/2,M/2]}}^{N/2} \!\!\!\! h_\mathrm{a}\left(\alpha'; D_\phi\right) \times h \left( \alpha; \phi_\mathrm{up}+2\uppi \alpha'\cdot \mathrm{Id}, \phi_\mathrm{down} \right) \nonumber \\
           & + \sum_{\alpha'_x =-M/2}^{M/2}\sum_{\alpha'_y =-M/2}^{M/2} h_\mathrm{a}\left(\alpha'; D_\phi\right) \times h_\mathrm{c} \left( \alpha; \phi_\mathrm{up}+2\uppi \alpha'\cdot \mathrm{Id}, \phi_\mathrm{down} \right). \nonumber
        \end{align}
        
        Now, we can give a convolutive structure to the first double sum. Let us define $h_\mathrm{a}^0$ as 
        \begin{align}
            h_\mathrm{a}^0(\alpha') = \left\{ \begin{array}{l}
                 0 \text{ if } \alpha' \in [-M/2,M/2]\times[-M/2,M/2]  \\
                 h_\mathrm{a}(\alpha') \text{ otherwise.} 
            \end{array} \right.
        \end{align}
        
        Then, Equation~(\ref{eq:inter_approx_convol}) can be re-written as :
        \begin{align}
        & h_\mathrm{lec}(\alpha; \phi_\mathrm{up}, \phi_\mathrm{down}, D_\phi)  = \\
           &  \sum_{\alpha'_x =-N/2}^{N/2}\sum_{\alpha'_y =-N/2 }^{N/2} h_\mathrm{a}^0\left(\alpha'; D_\phi\right) \times h \left( \alpha; \phi_\mathrm{up}+2\uppi \alpha'\cdot \mathrm{Id}, \phi_\mathrm{down} \right) \nonumber \\
           & + \sum_{\alpha'_x =-M/2}^{M/2}\sum_{\alpha'_y =-M/2}^{M/2} h_\mathrm{a}\left(\alpha'; D_\phi\right) \times h_\mathrm{c} \left( \alpha; \phi_\mathrm{up}+2\uppi \alpha'\cdot \mathrm{Id}, \phi_\mathrm{down} \right), \nonumber
        \end{align}
        which amounts to a convolution in the first double sum :
        \begin{align}
        & h_\mathrm{lec}(\alpha; \phi_\mathrm{up}, \phi_\mathrm{down}, D_\phi)  = h_\mathrm{a}^0(D_\phi) \star h \left(\phi_\mathrm{up}, \phi_\mathrm{down} \right)(\alpha) \label{eq:approx_corono} \\
           & + \sum_{\alpha'_x =-M/2}^{M/2}\sum_{\alpha'_y =-M/2}^{M/2} h_\mathrm{a}\left(\alpha'; D_\phi\right) \times h_\mathrm{c} \left( \alpha; \phi_\mathrm{up}+2\uppi \alpha'\cdot \mathrm{Id}, \phi_\mathrm{down} \right). \nonumber
        \end{align}
        
\begin{figure}
            \centering
            \includegraphics[width=0.92\linewidth]{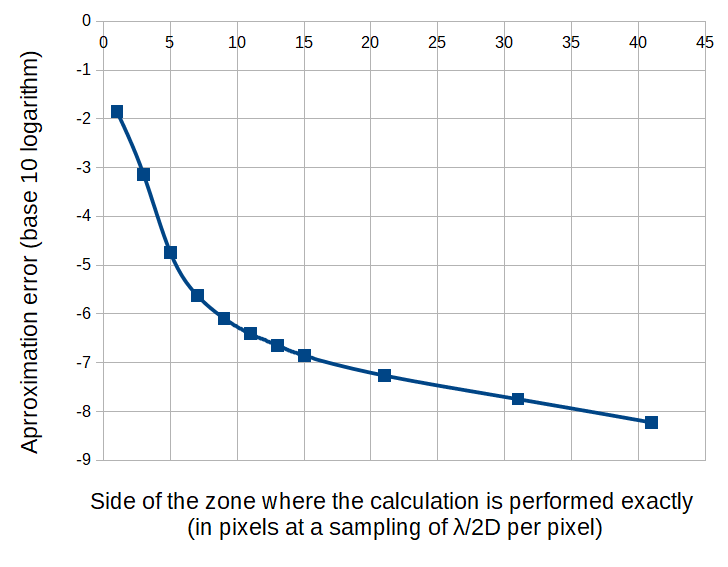}
            \caption{Evolution of the quality of the point spread function calculation as a function of the size of the zone where the influence of a Lyot coronagraph is taken into account}
            \label{fig:approx_corono}
        \end{figure}

        Since the cost of calculating the convolution product is negligible in comparison with the cost of calculating the double sum, the total cost of the calculation has diminished from $N^2$ times the cost of calculating $h_\mathrm{c}$ to $M^2$ times the cost of calculating $h_\mathrm{c}$. An important point is then to determine which is the quality of the approximation as a function of the side of the exact calculation zone, $M$. Figure~\ref{fig:approx_corono} presents the energy in the difference between the field calculated without approximation and the field calculated using Equation~(\ref{eq:approx_corono}). Various sizes $M$, varying from 1 to 41, are considered. The coronagraph that is considered here is a Lyot coronagraph of radius $2\lambda/D$. The root mean square of the upstream aberration is 100~nm at a wave-length $\lambda = 1,589$~nm, and the phase structure function is representative of SPHERE’s SAXO adaptive optics system. The normalisation factor is taken such that the coronagraphic point spread function without any aberration, without turbulence, has a total energy of 1. The sampling factor is chosen to satisfy exactly the Shannon--Nyquist condition, that is to say that a numeric field of $M$ pixels corresponds to an optic angular size of $M\lambda/2D$.
        
        There are two regimes of loss of precision due to the approximation. In the first regime, the approximation is crude because the size $M$ is insufficient, and the error decreases greatly with any increase in $M$. In the second regime, where $M>10$, the exact calculation is performed on a zone of extension $5\lambda/D$. Beyond this zone, the error committed by approximating a tilted coronagraphic point spread function by a non-coronagraphic one becomes negligible. Indeed, even if there is a Lyot (or Roddier \& Roddier) coronagraph, a light beam tilted by more than $5\lambda/2D$ will essentially not be modified by the coronagraph. Consequently, in the second regime, the approximation error is very low, and decreases more slowly.
        
        In conclusion, for coronagraphs such as the Lyot coronagraph or the Roddier \& Roddier coronagraph, and in particular the popular APLC, the cost of a long-exposure coronagraphic image simulation can be about a hundredfold the cost of a short-exposure coronagraphic image. \corr{In practice, the simulation of a $1024\times1024$-pixel long-exposure coronagraphic image results in a calculation time of less than five seconds on a single core of an office computer equipped with a 2.6 GHz Intel processor.} In principle, the approximation could be adapted to a four-quadrant phase mask coronagraph, where the convolutive approximation could be done far from the transitions.

    \subsection{Calculating the gradients}       
        The COFFEE criterion given by Equation~(\ref{eq:coffee}) can be separated into two parts.
        The gradient of the regularisation part, $\mathcal{R}$, is \corr{easier to calculate}, for the inclusion of turbulence into the model does not change it. As can be found in~\cite{paul_mesure_2014}  
        \begin{equation}
            \dfrac{\partial \mathcal{R}}{\partial \phi}(\phi) = - \dfrac{1}{\sigma^2_{\nabla \phi}} \Deltaup \phi \quad \text{ for } \quad \mathcal{R}(\phi) = \dfrac{1}{2\sigma^2_{\nabla \phi}}\sum_{x,y}\left\|\nabla \phi \right\|^2(x,y).
        \end{equation}
        
        Let us define $\mathcal{D}$ as the other part of the criterion, that is, the distance between the model $m$ and the data $i$ :
        \begin{equation}
            \mathcal{D} = \sum_{k,x,y} \dfrac{\left\| i(k,x,y) - m(\phi_\mathrm{up},\phi_\mathrm{down},k,x,y) \right\|^2}{2 \sigma^2(k,x,y)}.
        \end{equation}
        Then, using the complete expression of $m$, $\mathcal{D}$ is 
        \begin{equation}
            \mathcal{D} = \sum_{k,x,y} \dfrac{\left\| i(k,x,y) - f_k\times\left[h_\mathrm{det} \star h_\mathrm{lec}(\phi_\mathrm{up}+\phi_k,\phi_\mathrm{down})\right] (x,y) - b_k
 \right\|^2}{2 \sigma^2(k,x,y)},
        \end{equation}
        where 
        \begin{align}
            h_\mathrm{lec}&(\alpha; \phi_\mathrm{up}, \phi_\mathrm{down}, D_\phi)  =  \\
            & \sum_{\alpha'_x =-N/2}^{N/2}\sum_{\alpha'_y =-N/2}^{N/2} h_\mathrm{a}\left(\alpha'; D_\phi\right) \times h_\mathrm{c} \left( \alpha; \phi_\mathrm{up}+2\uppi \alpha'\cdot \mathrm{Id}, \phi_\mathrm{down} \right). \nonumber
        \end{align}
        
        The gradient of $\mathcal{D}$ with respect to $\phi$ can be calculated using the long-exposure coronagraphic point spread function as an intermediate variable :
        \begin{equation}
            \dfrac{\partial \mathcal{D}}{\partial \phi} = \sum_l\sum_m \dfrac{\partial \mathcal{D}}{\partial h_\mathrm{lec}(l,m)} \dfrac{\partial h_\mathrm{lec}(l,m)}{\partial \phi}.
        \end{equation}
        By using the analytic expression for $h_\mathrm{lec}$, this gradient is also 
        \begin{align}
            &\dfrac{\partial \mathcal{D}}{\partial \phi} \nonumber \\
            &= \sum_l\sum_m \dfrac{\partial \mathcal{D}}{\partial h_\mathrm{lec}(l,m)}\\
            &\;  \times \dfrac{\partial }{\partial \phi} \sum_{\alpha'_x =-N/2}^{N/2}\sum_{\alpha'_y =-N/2}^{N/2} \!\! h_\mathrm{a}\left(\alpha'; D_\phi\right) \times h_\mathrm{c} \left( l,m; \phi_\mathrm{up}+2\uppi \alpha'\cdot \mathrm{Id}, \phi_\mathrm{down} \right). \nonumber
        \end{align}
        Now, a rearrangement of the summation operators leads to
        \begin{align}
            \dfrac{\partial \mathcal{D}}{\partial \phi} & = \sum_{\alpha'_x =-N/2}^{N/2}\sum_{\alpha'_y =-N/2}^{N/2} h_\mathrm{a}\left(\alpha'; D_\phi\right) \\
              & \quad \times \sum_l\sum_m \dfrac{\partial \mathcal{D}}{\partial h_\mathrm{lec}(l,m)} \times \dfrac{\partial }{\partial \phi}  h_\mathrm{c} \left( l,m; \phi_\mathrm{up}+2\uppi \alpha'\cdot \mathrm{Id}, \phi_\mathrm{down} \right). \nonumber
        \end{align}
        
        One can note that the second line of this last expression is the gradient of the non-regularised criterion in the absence of turbulence, whose expression is given in the appendix of \cite{paul_high-order_2013}. Consequently, the structure of the calculation of this gradient is similar to the calculation of the point spread function itself. Thus, its calculation can also benefit from the acceleration described in the previous subsection.
        
        Now that the formalism and implementation of coronagraphic phase diversity in the presence of turbulence are described, let us move on to a numerical study of the robustness of the method.
        
\section{Numerical study of the robustness of the method}
\label{sec:sensitivity}
    In this section, we perform numerical simulations to study the impact on the quality of the COFFEE reconstruction of various discrepancies between the model $m$ and the actual imaging process leading to $i$. A simulation result when there is no discrepancy is presented in \cite{herscovici-schiller_analytic_2017}. We consider successive discrepancies in order to obtain a realistic simulation of the performance that can be expected of the technique on a real instrument.  
    
    \subsection{Parameters of the simulations}
        In all this section, the phase aberrations are expressed at a wavelength $\lambda = 1.589$~nm. \corr{The simulation is purely monochromatic, the wavelength being $\lambda = 1.589$~nm.}. The input parameters of the simulations include a phase structure function $D_\phi$ that is representative of turbulence at Paranal after correction by SPHERE’s adaptive optics SAXO. The upstream phase aberration, $\phi_\mathrm{up}$, has a root mean square of 50~nm, with an energy spectral density that follows a $f^{-2}$ statistics. This upstream phase aberration is what we wish to estimate thanks to COFFEE in the simulations. The downstream phase aberration, $\phi_\mathrm{down}$, has a root mean square of 20~nm, with the same energy spectral density as the upstream aberration. The coronagraph is an unapodized Lyot coronagraph, with a ratio of 95~\% between the diameter of the entrance pupil and the diameter of the Lyot stop. There are no amplitude aberrations in the system. 
        The total flux of the source is taken at $10^9$ photons, and the root mean square of the electronic noise of the detector is one electron per pixel. The diversity phase is supposed to be perfectly known. It is taken as a pure defocus, with root mean square 125~nm. The images in the focal plane have $128\times128$ pixels. The value of the sampling is chosen as 2, so the Shannon--Nyquist condition is respected, and the estimated phases are sampled over $64\times64$ pixels. 
        
    \subsection{Sensitivity to an error on the phase structure function}
        \corr{The analytic model that we use in the reconstruction algorithm requires the knowledge of the phase structure function of the post-adaptive optics residuals.}
        
        During operations, the statistics of the adaptive optics-corrected turbulence can be estimated using several techniques~\citep{veran97,sauvage_coronagraphic_2012}. A first family is telemetry techniques that give access to such data as the seeing or the wind speed. Those data can then be used jointly with the parameters of the adaptive optics system in a numeric simulation whose output is the phase structure function $D_\phi$. A second family of techniques is to use such real-time measurements of the adaptive optics system as residual slopes and command voltages. \corr{Since this section is concerned with performing simulations, we used the first option to produce the phase structure function.}
        
        Whether one or the other technique is chosen, the phase structure function will not be perfectly known in practice. In this subsection, we test how an error on the parameter $D_\phi$ impacts the quality of the reconstruction. We generate data using the parameters detailed in the previous subsections. We then perform COFFEE reconstructions using different phase structure functions, which are obtained by multiplying the “true” phase structure function $D_\phi$ by a factor $1-p$. For the sake of legibility, a reconstruction performed using $(1-p)\times D_\phi$ as phase structure function is called a reconstruction with a error of magnitude $p$ on $D_\phi$. This corresponds to a reconstruction where the seeing is underestimated, but with a correct knowledge of the wind velocity and the magnitude of the source. 
        \corr{Indeed, $D_\phi$ translates the phase residual statistics after the AO system, and this residual behaves in direct relation with the Fried parameter $r_0$~\citep{rigaut1998,fetick2018}, as the adaptive optics loop acts mainly as a filter. A more detailed analysis should also indicate the impact of a wind difference (hence only impacting the on-axis residuals), or a boiling effect, or mis-registration evolution, or the impact of additionnal dead actuators; here we perform a principle demonstration taking only $r_0$, which is the main contributor to the final correction error of the adaptive optics loop, hence producing a scaling on $D_\phi$.}

\begin{figure}
            \centering
            \includegraphics[width=0.8\linewidth]{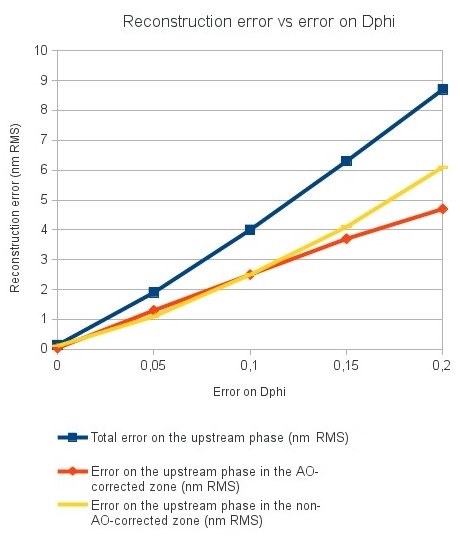}
            \caption{Estimation error as a function of the error on the knowledge of $D_\phi$. The total error is separated as an error in the zone that is corrected by the deformable mirror and the uncorrected zone.}
            \label{fig:erreur_Dphi}
        \end{figure}

        Figure~\ref{fig:erreur_Dphi} displays the root mean square error between the estimated upstream phase, $\widehat{\phi}_\mathrm{up}$, and the true upstream phase, $\phi_\mathrm{up}$ as a function on the error on $D_\phi$, for five different values of the error $p$. \corr{The error evolves in the same way for $p<0$.}. On SAXO, the adaptive optics of SPHERE, one can expect an error ranging from $5\%$ to $15\%$. Consequently, one can expect an error of one to three nanometres on the part of the aberrations that can be corrected by the deformable mirror. Thus, COFFEE is expected be a good candidate to measure and correct the quasi-static aberrations several times per night, using on-sky measurements.
        
    \subsection{Sensitivity to the noise level}
\begin{figure}
            \centering
           \includegraphics[width=1\linewidth]{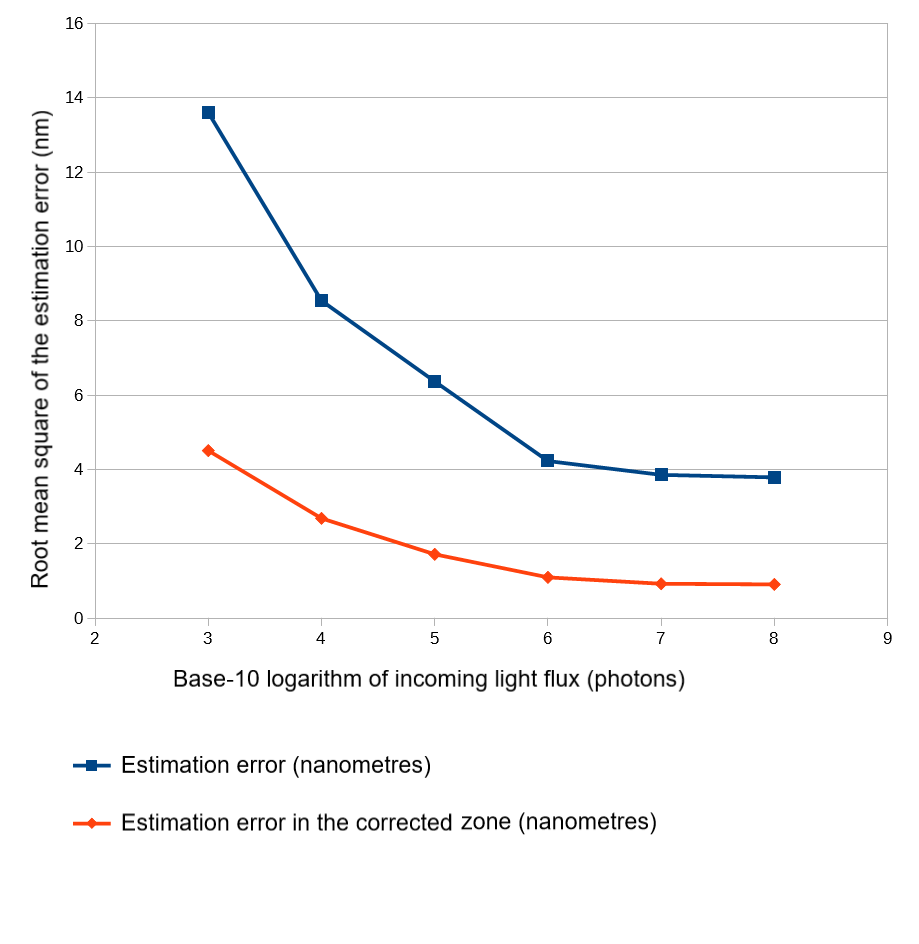}
            \caption{Root mean square of the estimation error of the upstream phase as a function of the incoming light flux (in photons). The estimation is performed with a 10~\% error on the phase structure function.}
            \label{fig:erreur_bruit}
        \end{figure}

        The incoming photonic flux has a critical impact on the noise level in the images. In this subsection, we study the impact of the photonic flux on the estimation quality of $\phi_\mathrm{up}$. We perform this study with a 10~\% error level on $D_\phi$, for we do not have proof that the various causes of estimation errors are independent. 
        When the light flux increases, the noise decreases, so we expect the quality of the estimation to increase. 
        This expectation is confirmed by the results shown on Fig.~\ref{fig:erreur_bruit}. It shows that the estimation error decreases with an increase in the flux. 
        Moreover, it shows that the quality of the estimation is much less sensitive to an increase in the incoming light flux at around one million incoming photons in the pupil.
        
        We can explain the order of magnitude of this critical point in a simple manner. The coronagraph blocks about $90\%$ of the incoming $10^6$ photons. About $10^5$ reach the $128\times128 = 16,384$~pixels of the detector, which amounts to an average of about 6 photons per pixel. Since the electronic noise is one electron per pixel, we deduce that the effect of an increase in the total flux is reduced if the average flux per pixel is such that the signal to noise ratio is higher than 5, which is easily reachable in a reasonable amount of time for a H magnitude of nine to twelve.
        
        Let us consider the case of an observation by the Very Large Telescope of a star of magnitude 15 in the visible. The photonic flux on the detector without a coronagraph is about $2\times 10^5$ photons per second in H-band. Since the data that we aim to use are typically exposures of a several hundred seconds, the resulting number of photons is in the range of $10^7$, which is quite enough to avoid the estimation to be limited by the noise level.

    \subsection{Sensitivity to the presence of a planet in the data}
        The data formation model that we use in COFFEE relies on the light propagation from a point source. However, for practical on-sky implementations, COFFEE must be able to estimate aberrations while observing planets. Here, we present COFFEE estimates performed on simulated images where there is a planet. We chose realistic parameters: we kept a $10\%$ error level on $D_\phi$, and an incoming light flux of $10^6$ photons in the pupil. The position of the planet is at an angle $3.5\lambda/D$ from the star. We simulated planets with flux ratios of $10^{-3}$, $10^{-4}$, and $10^{-5}$ between the star and the planet, so that we could test the various impacts that various planet fluxes could have on the estimation error.   
        
\begin{figure}
            \centering
            \includegraphics[width=\linewidth]{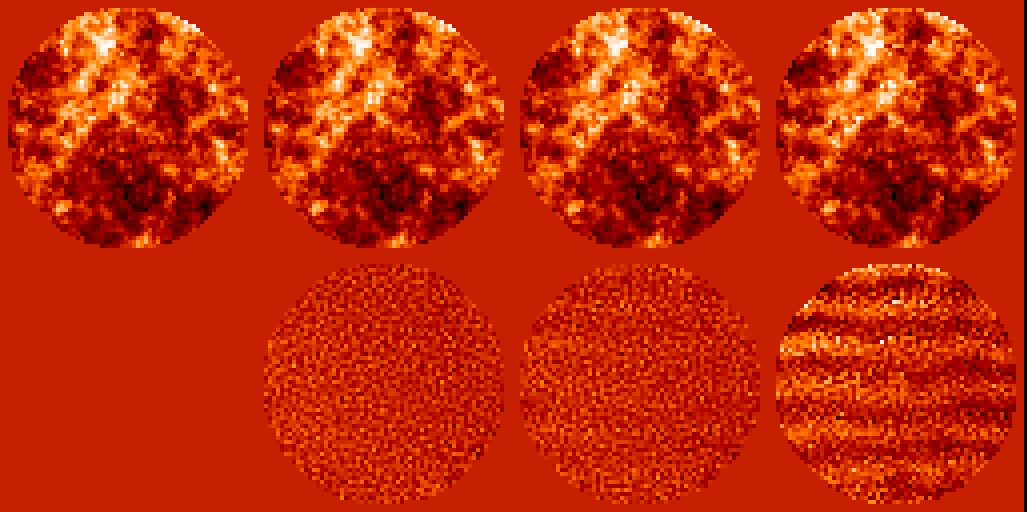}
            \caption{Top, from left to right: COFFEE estimate without a planet, COFFEE estimate with a planet of flux $10^{-5}$ relative to the star, COFFEE estimate with a planet of flux $10^{-4}$ relative to the star, and COFFEE estimate with a planet of flux $10^{-3}$ relative to the star. Bottom: difference between top and COFFEE estimate without a planet.}
            \label{fig:impact_planete}
        \end{figure}

        Figure~\ref{fig:impact_planete} presents COFFEE estimates with no planet in the data, and with planets whose fluxes are $10^{-5}$, $10^{-4}$, and $10^{-3}$ with respect to the star.
        
        In the case where the flux ratio is $10^{-5}$, the phase estimation is almost unperturbed by the planet: the root mean square of the difference between the reconstruction with a planet and the reconstruction without planet is only 0.35~nm. This difference is mainly due to the fact that the realisations of the noise in the data are different. 
        
        In the case where the flux ratio is $10^{-3}$, the phase estimation is strongly perturbed by the planet. In that case, the root mean square of the difference between the reconstruction with a planet and the reconstruction without planet is 6.0~nm. The error is mainly a sinusoidal structure whose image through the data formation model generates a strong speckle that looks like a planet.
        
        In the intermediate where the flux ratio is $10^{-4}$, the phase estimation is slightly perturbed by the planet. In that case, the root mean square of the difference between the reconstruction with a planet and the reconstruction without planet is 0.5~nm. 
        
\begin{figure}
            \centering
            \includegraphics[width=\linewidth]{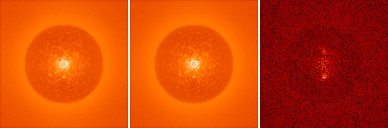}
            \caption{Left: focused image taken as a COFFEE input with the presence of a planet. The planet is at a third of the adaptive optics correction radius, below the star. The flux of the planet is $10^{-3}$ times the flux of the star. Middle: direct model corresponding to the COFFEE reconstruction. Right: difference between direct model corresponding to the COFFEE reconstruction (middle) and direct model corresponding to the true aberration; the impact of the planet on the estimation quality is clearly visible.}
            \label{fig:diff_flux_3}
        \end{figure}

\begin{figure}
            \centering
            \includegraphics[width=\linewidth]{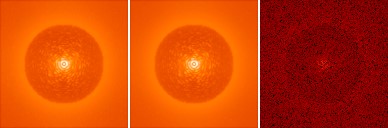}
            \caption{Same as Fig.~\protect\ref{fig:diff_flux_3}, except this time the flux of the planet is $10^{-5}$ times the flux of the star. Contrarily to Fig.~\protect\ref{fig:diff_flux_3}, the planet is invisible in the data, and it has no impact on the estimation quality.}
            \label{fig:diff_flux_5}
        \end{figure}

        These results show that COFFEE may or may not mistake a planet in the data for a speckle, depending on its light flux. This influence of the light flux of the planet can be explained by comparing it with the regularisation level. If the flux of the planet is high, the associated noise in the pixels that image it is low, and the planet then has an important impact on the criterion, leading to a sinusoidal estimation for $\phi_\mathrm{up}$. On the opposite, if the flux of the planet is low, the associated noise in the pixels that image it is high, and the planed then has a low impact in the criterion, so the regularisation will prevent the emergence of an artefact. This is especially obvious if one compares Figs~\ref{fig:diff_flux_3} and\ref{fig:diff_flux_5}.
        
    We conclude that, for a planet whose light flux compared to its star is less than $10^{-4}$, its presence in the data will account for an estimation error of root mean square less than half a nanometre in the SPHERE-like case of a fifty-nanometre upstream aberration. If, however, a planet of flux higher than $10^{-4}$ happened to be in the field, then it would be clearly visible in the raw data (compare Figs~\ref{fig:diff_flux_3} and \ref{fig:diff_flux_5}). In that case, one could choose to ignore a small region about the planet in the COFFEE input data. The alternative is to implement a model of the planet as a source point in the COFFEE algorithm. Even in the presence of an imperfect knowledge of the phase structure function, and with a limited exposure time, the expected error root mean square of the estimation error is of the order of a nanometre in the adaptive optics-corrected zone, for a 50-nanometres root mean square phase aberration. With this encouraging simulation result in mind, we proceed to the laboratory validation.
    

\section{Laboratory validation of the coronagraphic phase diversity in the presence of residual turbulence}
\label{sec:lab}
    \subsection{Strategy of validation}
        \subsubsection{Aim}
            The aim of our experiment was to use the coronagraphic phase diversity to estimate a static phase aberration upstream of a coronagraph by using post-coronagraphic images as input data. The experiment was performed in a controlled environment, the MITHiC testbed.
        \subsubsection{The MITHiC testbed}
\begin{figure}
                \centering
                \includegraphics[width=\linewidth]{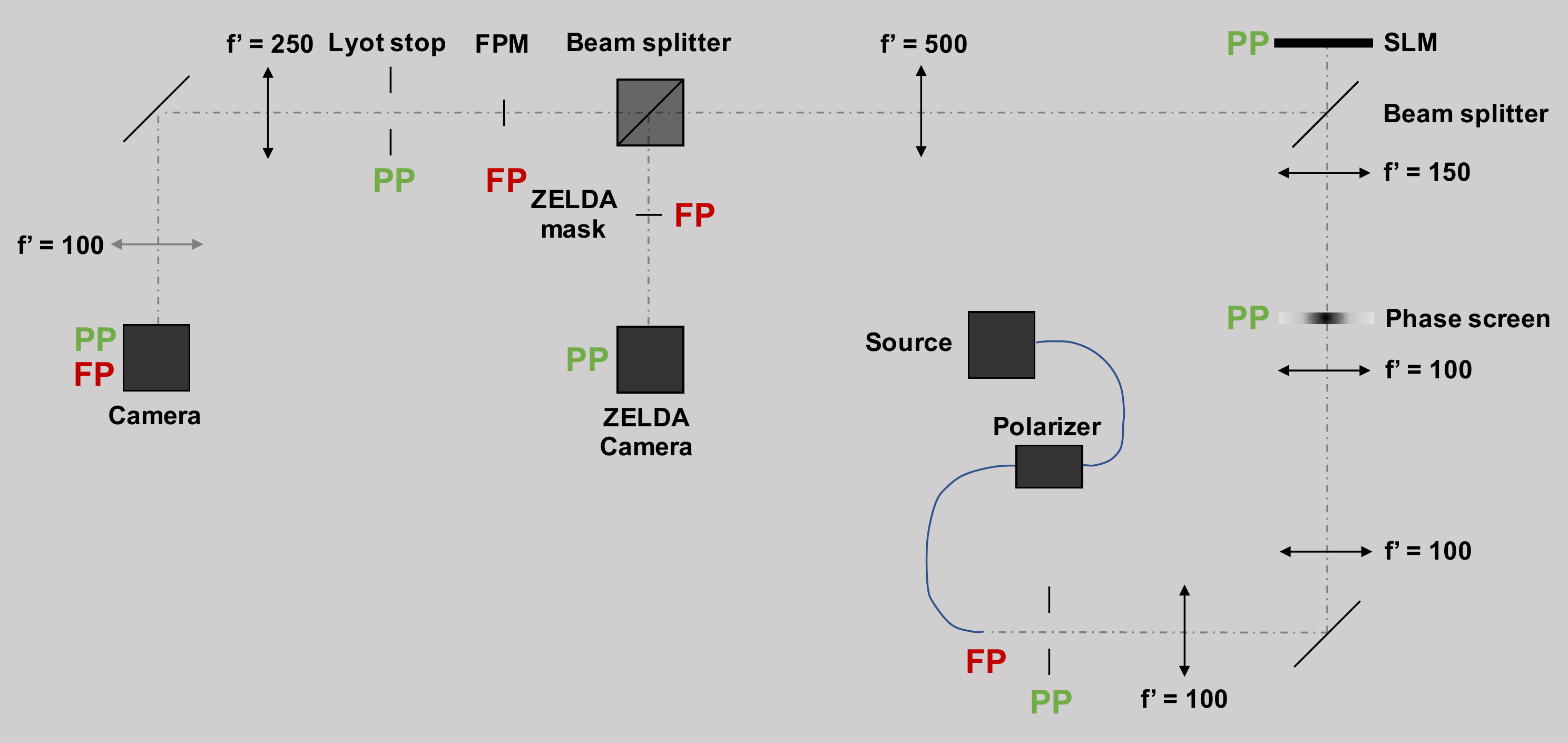}
                \caption{Schematical representation of MITHiC, reproduced from \protect\cite{leboulleux_optimal_2018}. Courtesy Lucie Leboulleux.}
                \label{fig:schema_mithic}
            \end{figure}

            MITHiC is the \textit{Marseille Imaging Testbed for High Contrast imaging}. It has been developed at the laboratoire d’astrophysique de Marseille (LAM) for almost ten years~\cite{ndiaye_lab_2012}. It is schematically described on Fig.~\ref{fig:schema_mithic}. 
            
            The light source is a super-luminescent diode. It emits light in a narrow spectral band centred around $\lambda= 677$~nm. In all our data processing, it is considered as a monochromatic light source. The light undergoes linear polarisation, and injected on the bench through an entrance pupil. 
            
            It is propagated in a second pupil plane, where it passes through a rotating transparent phase screen. On this screen are engraved random path differences whose statistics is that of atmospheric turbulence corrected by SAXO, the adaptive optics system of the SPHERE instrument. The scale of these path differences is such that the \corr{45-nm-RMS} phase shift that they create at the working wavelength of 677~nm is the same as the  \corr{100-nm-RMS} phase shift created by the SAXO-corrected atmospheric turbulence at 1,600~nm, which is a typical observational wavelength for SPHERE. \corr{The phase screen used on MITHIC to produce the wavefront errors (both the residual turbulence, or dedicated static patterns) have been specified by LAM and realized by SILIOS company on a pixel map interface. LAM has provided exactly the phase map (pixel by pixel depth graduated in nanometers) to be engraved on the phase screen. The realization of SILIOS has been checked at LAM with a high-resolution ZYGO interferometer after delivery, and they are correct at a nanometric level, which guarantees that the statistic, as well as the power law and RMS across the aperture are the expected ones.}
            
            In a third pupil plane, the light meets the surface of a spatial light modulator. The presence of this element is the reason why the light is polarised in the first place. The spatial light modulator is used as a high resolution deformable mirror.
            
            The light is then split into two paths using a beam splitter. The auxiliary path, or ZELDA path~\citep{ndiaye_calibration_2013}, can be used to perform ZELDA experiments, or to use a HASO wavefront sensor as a calibration reference, which is what we did. The main path, which is the one of interest for us here, comprises a Roddier \& Roddier focal plane mask, and a Lyot stop in the next pupil plane. 
            
            Finally, the light reaches a focal plane camera. This focal plane can be turned into a pupil plane by introducing a movable lens in the light beam.
            
        \subsubsection{Validation strategy}
            Our goal was to validate aberration estimation using coronagraphic phase diversity in the presence of residual adaptive optics-corrected turbulence. The plainest strategy imaginable would have been to introduce a known pupil-plane phase aberration upstream of the coronagraph, take focused and diversity images, process them with COFFEE, and compare the output of COFFEE to the known aberration. 
            However, this plain strategy would have needed that the optical testbed be absolutely perfect, that is to say that the introduced aberration be perfectly known. Since aberrations always exist on the bench, and COFFEE is an absolute wavefront sensor, this was not feasible. So we proceeded in two times, using a differential estimation strategy, as for example in \cite{herscovici2018experimental}.
            
            First, we did not introduce any aberration. Let us call $\phi_\mathrm{up}^0$ the static aberration on the bench. We took a focused image and a diversity image, which we collectively denote by $\mathbf{i^0}$. We used $\mathbf{i^0}$ as an input in COFFEE, the output being our estimate of the static aberration on the bench, $\widehat{\phi_\mathrm{up}^0}$.
            
            Then, we used the spatial light modulator to introduce a known aberration, $\phi_\mathrm{up}^F$. This did not suppress the aberration $\phi_\mathrm{up}^0$ on the bench, so the resulting total aberration on the bench was $\phi_\mathrm{up}^0 + \phi_\mathrm{up}^F$. We took a focused image and a diversity image, which we collectively denote by $\mathbf{i^1}$. We used $\mathbf{i^1}$ as an input in COFFEE, the output being our estimate of the total aberration on the bench, $\widehat{\phi_\mathrm{up}^0 + \phi_\mathrm{up}^F}$.
            
\begin{figure}
                \centering
               \includegraphics[width=\linewidth]{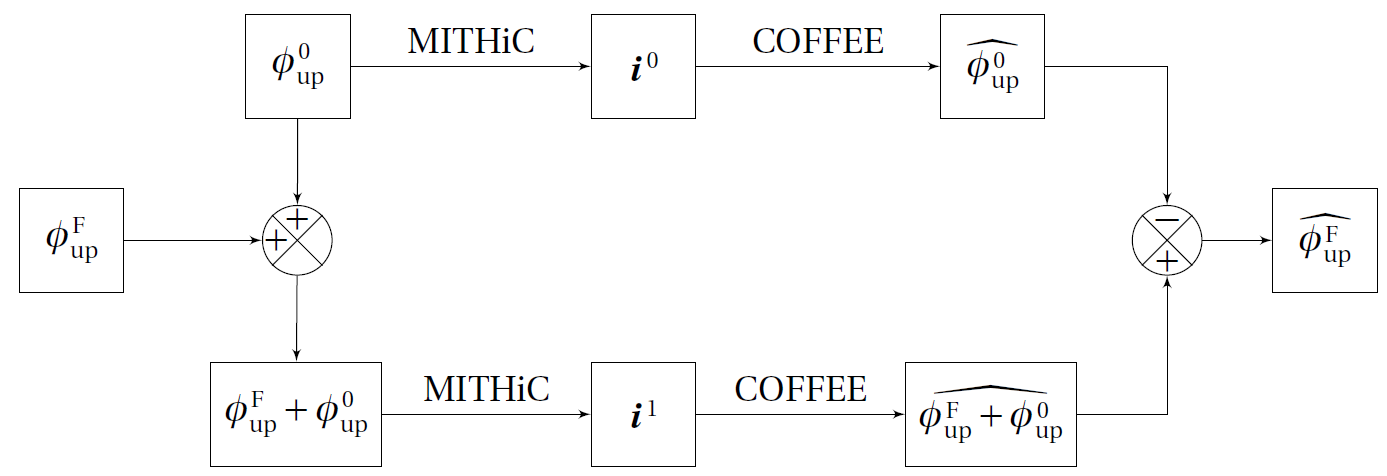}
                \caption{Strategy for the experimental validation of coronagraphic phase diversity in the presence of residual turbulence.}
                \label{fig:strategie_validation}
            \end{figure}

            The last step is to compute the difference $\widehat{\phi_\mathrm{up}^0 + \phi_\mathrm{up}^F} - \widehat{\phi_\mathrm{up}^0}$, which we use as the estimate $\widehat{\phi_\mathrm{up}^F}$ of $\phi_\mathrm{up}^F$. 
            This validation strategy is described symbolically on Fig.~\ref{fig:strategie_validation}. 
            
    \subsection{Calibration of the parameters of the optical model}
        Since COFFEE relies on a model of image formation, some calibrations are necessary to perform an estimation. We detail these calibrations here.
        \subsubsection{Sampling on the detector}
\begin{figure}
                \centering
                \includegraphics[width=\linewidth]{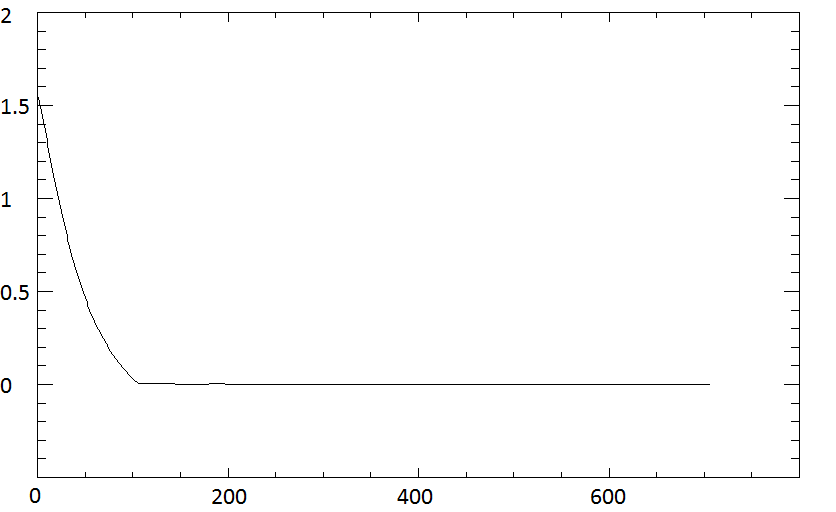}
                \caption{Circular average of the modulation transfer function of MITHiC}
                \label{fig:ftm}
            \end{figure}

            To determine the sampling on the detector, we took a non-coronagraphic image whose size is $1,000\times1,000$~pixels. The modulus of its Fourier transform is the modulation transfer function. This modulation transfer function, whose circular average is shown on Fig.~\ref{fig:ftm}, goes to zero at spatial frequency pixel number 107. Consequently, the sampling factor $sf$ on the detector is
            \begin{equation}
                sf = \frac{1000}{107\pm1} = 9.43\pm0.1 .
            \end{equation}
            
            In order to shorten calculation times, the data used as input to COFFEE were under-sampled by a factor 4. The resulting sampling being higher than 2, the Shannon--Nyquist condition is satisfied, so there is no loss of information in the reconstruction. 
        \subsubsection{Lyot ratio}
            The Lyot ratio is defined as the ratio between the diameter of the entrance pupil and the diameter of the Lyot stop. It is a necessary parameter for the direct model. In order to determine it, we introduced the Lyot stop --- but not the Roddier \& Roddier focal mask --- on the bench. Then, just as for the determination of the sampling on the detector, we took an image and examined the cut of the corresponding modulation transfer function. It is located between pixels 99 and 100. Since the Lyot ratio is proportionnal to the ratio of the cut frequencies, the Lyot ratio is 
            \begin{equation}
                r_\mathrm{L} = \frac{99.5\pm0.5}{107\pm1}=0.93\pm0.01 .
            \end{equation}
            
        \subsubsection{Detector noise}
            The only information on the electronic noise of the camera given by the manufacturer is that its root mean square is less than 5 electrons per pixels~\cite{photometrics_coolsnaptm_2014}. A previous calibration had found a root mean square value of one electron per pixel. We calculated the root mean square of a stack of 4000 images taken in complete darkness, which yielded a root mean square of 1.6 electrons per pixel. We use this value $\sigma_\text{det}$ as the detector noise in the denominator in Equation~\ref{eq:coffee}. The exact formula for the denominator is $\sigma^2 = \sigma^2_\text{photon}+\sigma^2_\text{det}$, where $\sigma^2_\text{photon}$ is directly estimated from the images~\citep{mugnier2004mistral}.
        
        \subsubsection{Phase structure function}
\begin{figure}
                \centering
                \includegraphics[width=0.6\linewidth]{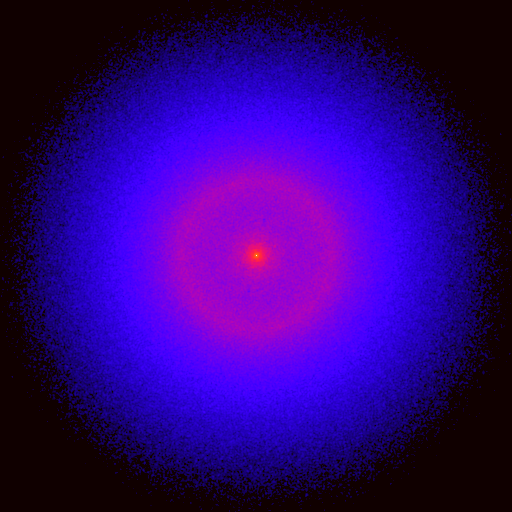}
                
                \includegraphics[width=\linewidth]{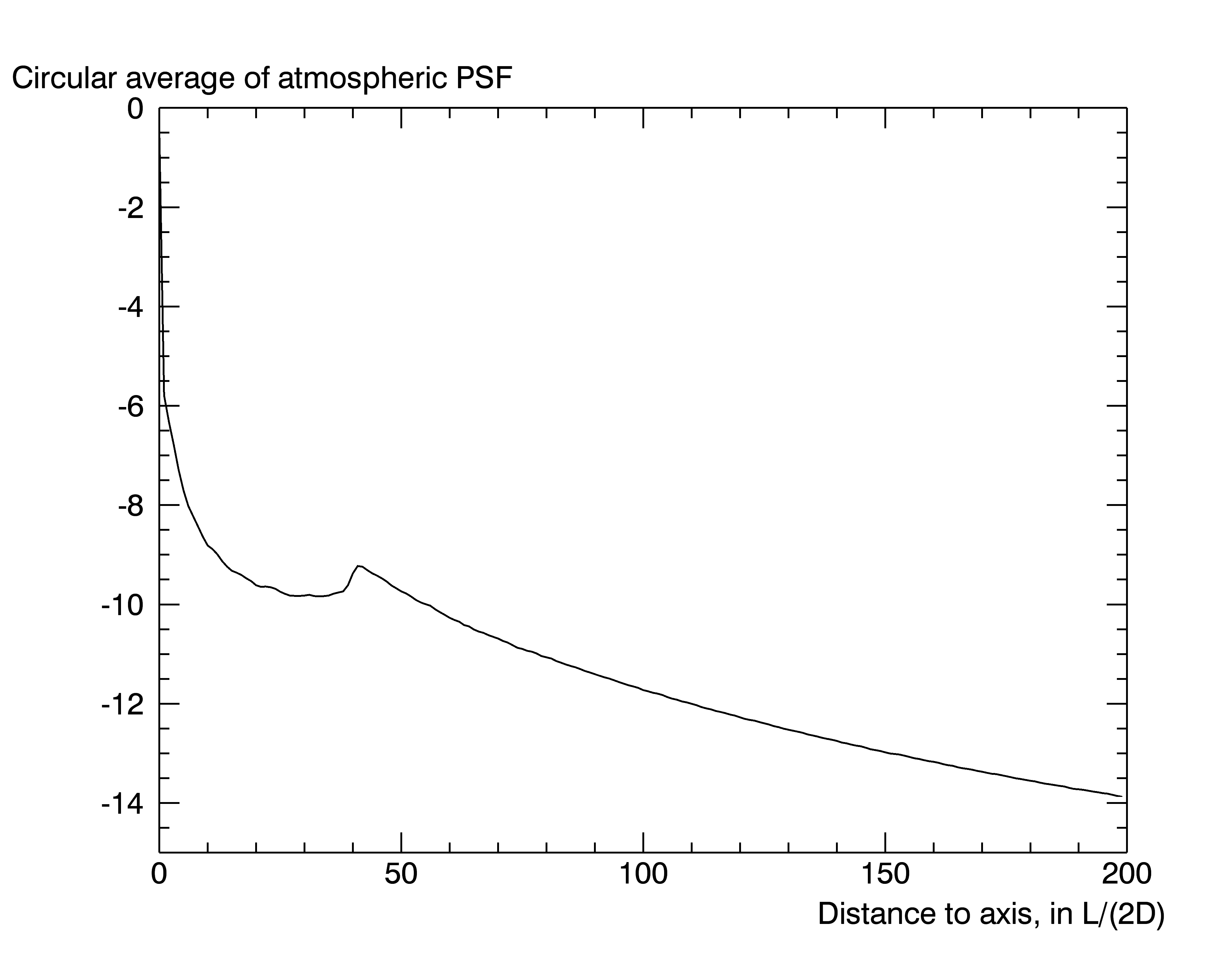}
                \caption{The atmospheric point spread function $\left|\mathcal{F}\left[\exp\left(-D_\phi/2\right)\right]\right|$ used in the experimental estimations \corr{(top) and its circular average (bottom, log scale).}}
                \label{fig:dphi}
            \end{figure}

            A key parameter of the direct model in the presence of turbulence is the phase structure function of the post-adaptive optics turbulence. In order to estimate it in our case, we took the specification file of the rotating phase screen, and calculated a variance over as many distinct realisation as there were non-overlapping disks in the adaptive-optics corrected turbulence path difference strip on the phase screen. The absolute value of the corresponding atmospheric point spread function \corr{(defined by Eq.~\ref{eq:ha})}, $\left|\mathcal{F}\left[\exp\left(-D_\phi/2\right)\right]\right|$, is displayed on Fig.~\ref{fig:dphi}. \corr{The correction limit of the adaptive optics at a radius of 20 \(\lambda/D\) is clearly visible.}
        \subsubsection{Uncertainty on the introduced diversity}
\begin{figure}
                \centering
                \includegraphics[width=\linewidth]{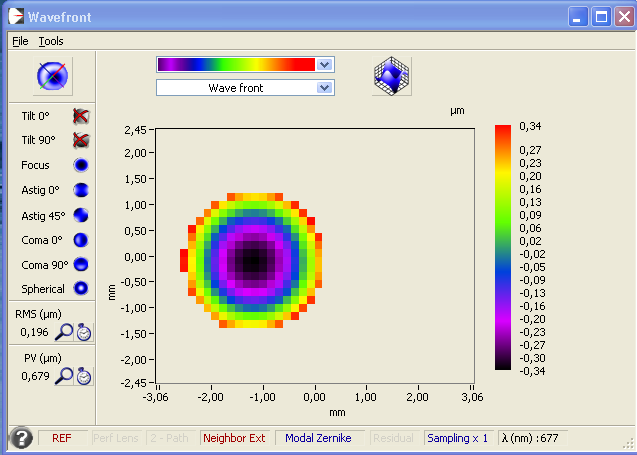}
                \includegraphics[width=\linewidth]{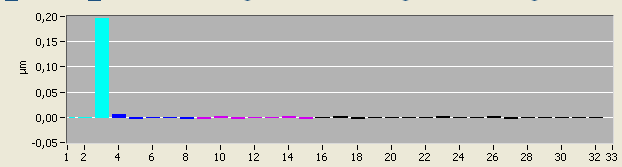}
                \caption{Top: HASO measurement for a 200 nm root mean square pure defocus control applied on the spatial light modulator. Bottom: Projection of this measurement on the 32 first Zernike modes.}
                \label{fig:haso}
            \end{figure}

            The diversity phase was introduced by the spatial light modulator. Since any error on the diversity phase impacts the quality of the estimation, it is important to check the exactness of the introduced diversity. In order to do so, we first introduced a command for a 200~nm defocus on the spatial light modulator. We measured the defocus that was effectively introduced thanks to an Imagine Optic HASO wavefront sensor. The HASO measurement, displayed on Fig.~\ref{fig:haso}, shows that the introduced phase diversity is really a pure defocus. When projected on the first thirty-two Zernike polynomials, the HASO measurement is of a 196~nm root mean square aberration, 195~nm of whose are concentrated in the defocus. This shows that the spatial light modulator produces phase is close to that which it is commanded to produce. 
            
\begin{figure}
                \centering
                \includegraphics[width=1.1\linewidth]{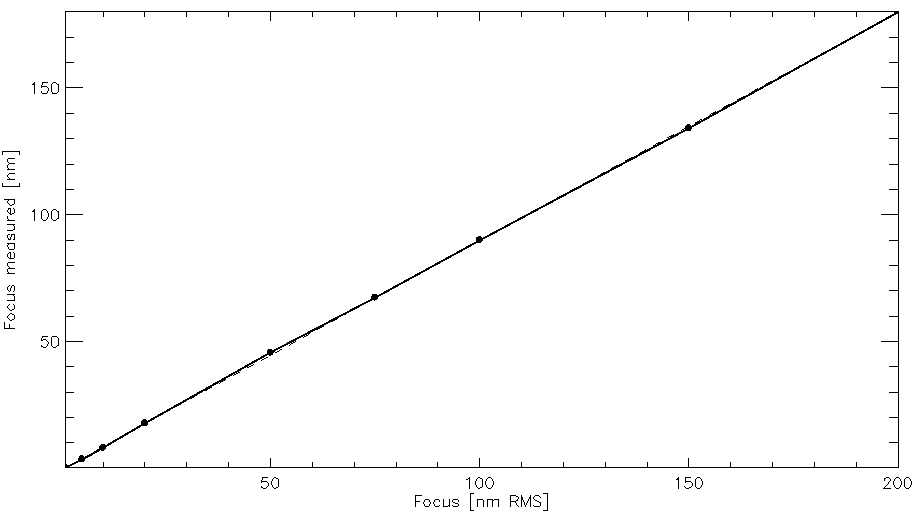}
                \caption{Response of the spatial light modulator to a defocus command, as measured by the HASO.}
                \label{fig:linearite}
            \end{figure}

            Another important matter is the linearity of the response of the spatial light modulator. We tested the amplitude of the defocus, as measured by the HASO, for various commands. Figure~\ref{fig:linearite} shows the excellent linearity of the response, with a linear correlation coefficient of 0.994. Consequently, for a defocus diversity whose root mean square is 75 nanometres, the error on the diversity should have an impact of less than a nanometre on the estimation~\citep{blanc2003calibration}.

    \subsection{Estimation of the reference wavefront}
        \subsubsection{Data}
            We aligned the coronagraph on the bench, activated the turbulence simulator, and turned on the light source. The spatial light modulator is sent a flat command. The resulting phase aberration is denoted by $\phi_\mathrm{up}^0$ on Fig.~\ref{fig:strategie_validation}. A stack of 400 images is saved. Each exposure lasts 0.035~s, with a wait of 0.020~s between exposures. After that, the light source is turned off, and a stack of 400 images is saved in the same condition. The first stack (the data stack) is then averaged; the median of each pixel of the second stack (the background stack) is calculated; and then the background is subtracted from the data sum average. The resulting data is then under-sampled by a factor 4 for the sake of calculation speed, and constitutes the data $\mathbf{i_\mathrm{foc}^0}$ that will be input into COFFEE.
            
\begin{figure}
                \centering
                \includegraphics[width=1.05\linewidth]{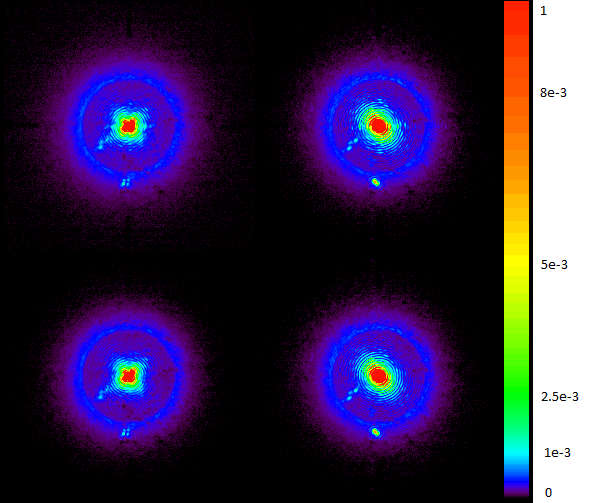}
                \caption{Bottom left: focused experimental coronagraphic image for the reference wavefront. Bottom right: diversity (75~nm defocus) experimental coronagraphic image for the reference wavefront. Top left: focused experimental coronagraphic image for the high-frequency aberration wavefront. Top right: diversity (75~nm defocus) experimental coronagraphic image for the high frequency aberration wavefront. The colorbar is expressed relatively to the maximum of intensity and is non-linear (asinh function) to best display the structure of the point spread functions and the differences between them. There is a slight difference between the top and the bottom row, for example there is a speckle on the horizontal axis just on the right of the core of the PSF in the high-frequency aberration images (top) that is absent from the reference wavefront images (bottom). }
                \label{fig:images_mithic}
            \end{figure}

            Then the whole procedure is repeated, this time with the spatial light modulator being sent a 75~nm defocus. The resulting data, $\mathbf{i_\mathrm{div}^0}$, joins $\mathbf{i_\mathrm{foc}^0}$ to constitute $\mathbf{i^0}$, the complete COFFEE input for the reconstruction.
            This input is displayed on the bottom part of Fig.~\ref{fig:images_mithic}.
            
        \subsubsection{Reconstruction}
\begin{figure}
                \centering
                \includegraphics[width=0.4\linewidth]{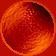}
                \caption{Estimated reference phase, $\widehat{\phi^0_\mathrm{up}}$}
                \label{fig:phase_ref}
            \end{figure}

            We perform an estimation of $\phi_\mathrm{up}^0$ using $\mathbf{i^0}$ as input data, along with the model of the bench calibrated in the previous section. 
            The estimate $\widehat{\phi^0_\mathrm{up}}$ of the reference wavefront $\phi^0_\mathrm{up}$ is shown on Fig.~\ref{fig:phase_ref}. Its root mean square is 24~nm.
            
    \subsection{Estimation of a high-frequency aberration}
        \subsubsection{Data}
\begin{figure}
                \centering
                \includegraphics[width=0.4\linewidth]{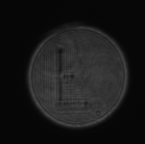}
                \caption{Non-coronagraphic pupil image of a high-frequency aberration.}
                \label{fig:f_pupille}
            \end{figure}

            The spatial light modulator is now sent a F-shape command. The corresponding pupil image (taken when no coronagraph was in place) is displayed on Fig.~\ref{fig:f_pupille}. The root mean square of this command is 11~nm.
            
            The resulting phase aberration is denoted by $\phi_\mathrm{up}^F$ on Fig.~\ref{fig:strategie_validation}. Just like before, a stack of 400 images is saved. Each exposure lasts 0.035~s, with a wait of 0.020~s between exposures. After that, the light source is turned off, and a stack of 400 images is saved in the same condition. The first stack (the data stack) is then averaged; the median of each pixel of the second stack (the background stack) is calculated; and then the background is subtracted from the data sum average. The resulting data is then under-sampled by a factor 4 for the sake of calculation speed, and constitutes the data $\mathbf{i_\mathrm{foc}^1}$ that will be input into COFFEE.
            
            Then the whole procedure is repeated, this time with the spatial light modulator being sent a 75~nm defocus on top of the F-shape aberration. The resulting data, $\mathbf{i_\mathrm{div}^1}$, joins $\mathbf{i_\mathrm{foc}^1}$ to constitute $\mathbf{i^1}$, the complete COFFEE input for the reconstruction.
            This input is displayed on the top part of Fig.~\ref{fig:images_mithic}.
            

        \subsubsection{Reconstruction}
\begin{figure}
                \centering
                \includegraphics[width=0.4\linewidth]{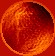}
                \caption{Estimated high-frequency aberration phase, $\widehat{\phi^F_\mathrm{up}}$}
                \label{fig:phase_f}
            \end{figure}

            We perform an estimation of $\phi_\mathrm{up}^0+\phi_\mathrm{up}^F$ using $\mathbf{i^1}$ as input data, along with the model of the bench calibrated in the previous section.
            The estimate $\widehat{\phi^0_\mathrm{up}+\phi_\mathrm{up}^F}$ of the \corr{high-frequency} wavefront $\phi^0_\mathrm{up}+\phi_\mathrm{up}^F$ is shown on Fig.~\ref{fig:phase_f}. Its root mean square is 28~nm.
            
        \subsubsection{Differential reconstruction}
\begin{figure}
                \centering
                \includegraphics[width=0.4\linewidth]{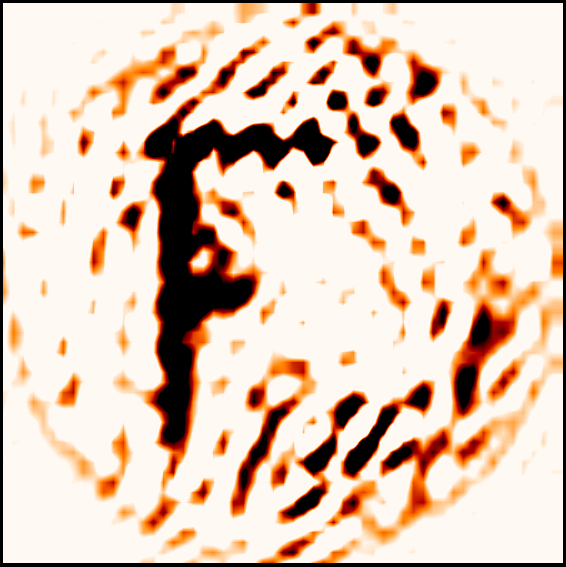}
                \caption{$\phi_\mathrm{up}^F$, the 13-nm root mean square estimate of the introduced 11-nm root mean square F-shape aberration.}
                \label{fig:F}
            \end{figure}

            The difference between $\widehat{\phi_\mathrm{up}^0+\phi_\mathrm{up}^F}$ and $\phi_\mathrm{up}^0$ is our estimate $\widehat{\phi_\mathrm{up}^F}$ of the introduced F-shape aberration, $\phi_\mathrm{up}^F$.
            It is displayed on Fig.~\ref{fig:F}. The estimate has a root mean square of 13 nanometres, whereas the introduced aberration has a root mean square of 11 nanometres. We conclude that we are able to reconstruct a high-frequency aberration of about 10 nanometres with a 2 nanometres accuracy, using the scientific camera of a coronagraphic system in the presence of turbulence.
            
            \corr{The total computing time necessary for the reconstruction of the aberration is about two hours and a half on a single core of an office computer equipped with a 2.6 GHz Intel processor, running a COFFEE program written in Interactive Data Language, without much optimisation. Therefore, using a modern computer running an optimised code, quasi-static aberrations compensation could be performed on an instrument such as SPHERE at least once per hour.}

\section{Conclusions}
    We have presented coronagraphic phase diversity through turbulence as a post-coronagraphic wavefront sensor adapted to high-precision wavefront measurement in the presence of adaptive optics-corrected turbulence. We have performed realistic simulations which show that the estimation error when used on a high-contrast system such as SPHERE should not exceed a few nanometres, whereas the quasi-static aberration of SPHERE is about 50 nanometres. Finally, we have performed a laboratory demonstration of the validity of the technique, reconstructing a high-frequency aberration with a precision of about two nanometres through a coronagraph and turbulence. Our team’s next step in to perform a measurement and correction of quasi-static aberrations on SPHERE, thanks to data collection by A. Vigan and M. N’Diaye. \corr{Since we demonstrated coronagraphic phase diversity in the presence of residual adaptive-optics-corrected turbulence, since there are no intrinsic chromatic limitations with COFFEE, and since the execution time of the program allows it to be executed several time during the night, we hope to soon prove that it is efficient on-sky. }
    
\section*{Acknowledgements}
The PhD work of O. Herscovici-Schiller was co-funded by
CNES and ONERA. We thank J.-M. Le Duigou for his support.
This work received funding from the E.U. under FP7
Grant Agreement No. 312430 OPTICON, from the CNRS
(Défi Imag'In), and from ONERA in the framework of the
VASCO research project. We thank Raphaël Galicher for helpful criticism of a first draft of this paper. 
We thank the reviewer for their careful review and constructive comments, which helped us to improve this article.




\bibliographystyle{mnras}
\bibliography{biblio} 


\bsp	
\label{lastpage}
\end{document}